\documentclass{PoS}
\PoS{PoS(LAT2005)070}
\usepackage[dvips]{graphicx}
\title{Lattice QCD Evidence for Exotic Tetraquark Resonance }
\ShortTitle{Lattice QCD Evidence for Exotic Tetraquark Resonance }

\author{\speaker{Hideo Suganuma} and Kyosuke Tsumura \\

Department of Physics, Kyoto University, 
Kitashirakawaoiwake, Sakyo, Kyoto 606-8502, Japan\\

         E-mail: \email{suganuma@scphys.kyoto-u.ac.jp}, 
                 \email{tmura@scphys.kyoto-u.ac.jp}}

\author{Noriyoshi Ishii \\

Department of Physics, Tokyo Institute of Technology,
Ohokayama, Tokyo 152-8551, Japan\\

         E-mail: \email{ishii@th.phys.titech.ac.jp}}

\author{Fumiko Okiharu \\

Department of Physics, Nihon University, 
Kanda-Surugadai, Chiyoda, Tokyo 101-8308, Japan\\
 
         E-mail: \email{oharu@phys.cst.nihon-u.ac.jp}}

\abstract{
We study the manifestly exotic tetraquark
D$_{\rm s0}^{++}(cu\bar s \bar d)$ and 
the scalar tetraquark $f_0(ud \bar u \bar d)$ 
in SU(3)$_c$ anisotropic quenched lattice QCD. 
We adopt the $O(a)$-improved Wilson (clover) 
fermion at $\beta=5.75$ on $12^3 \times 96$ with renormalized 
anisotropy $a_s/a_t=4$, and investigate the correlator of 
the four-quark (4Q) system, $cu \bar s \bar d$, 
with various quark masses including the idealized SU(4)$_f$ case.
For $f_0(ud\bar u \bar d)$ etc., we only consider connected diagrams at the quenched level, i.e., 
the tetraquark $f_0(ud\bar u\bar d)$ is identical with 
D$_{\rm s0}^{++} (cu\bar s \bar d)$ in the idealized SU(4)$_f$ case.
First, for comparison, we study the lowest $q\bar q$ scalar meson, and find that 
it has a large mass of about 1.37GeV after chiral extrapolation, 
which corresponds to $f_0(1370)$.
Second, we investigate the lowest 4Q state  
in the spatial periodic boundary condition, and find that 
it is just a scattering state of two pseudoscalar mesons, as is expected. 
Third, to extract spatially-localized 4Q resonances, 
we use the Hybrid Boundary Condition (HBC) method, 
where anti-periodic and periodic boundary conditions are imposed 
on quarks ($c$,$u$) and antiquarks ($\bar s$,$\bar d$), respectively. 
By applying the HBC on a finite-volume lattice, the threshold of the two-meson scattering state is 
raised up, while the mass of a compact 4Q resonance is almost unchanged.
In HBC, we find a nontrivial 4Q resonance state 
about 100 MeV below the two-meson threshold in some quark-mass region.
Its chiral behavior largely differs from a two-meson scattering state.
The scalar tetraquark $f_0(ud\bar u\bar d)$ is found to have the mass of 
about 1.1GeV after chiral extrapolation, and seems to correspond to $f_0(980)$.
Then, the manifestly exotic tetraquark D$_{\rm s0}^{++}(cu\bar s \bar d)$ 
would exist around 1GeV in the idealized SU(4)$_f$ chiral limit.
Finally, MEM analysis is applied to obtain the spectral function of the 4Q system.
}

\FullConference{XXIIIrd International Symposium on Lattice Field Theory\\

		 25-30 July 2005\\

		 Trinity College, Dublin, Ireland}

\begin{document}

\section{Introduction}

\vspace{-0.1cm}

The tetraquark [1-7] 
is an interesting subject in hadron physics. 
The tetraquark picture was first proposed for scalar mesons \cite{J77}.
Recently, charmed tetraquark candidates were discovered~\cite{Ds,X}. 

\vspace{-0.1cm}

\subsection{Experimental discoveries of tetraquark candidates}

In recent years, various candidates of multi-quark hadrons 
have been experimentally observed \cite{Ds,X} and 
theoretically studied [4-9].
For instance, D$_{\rm s0}^+$(2317) \cite{Ds}, D$_{\rm s1}^+$(2460), X(3872) \cite{X} and Y(4260) 
are expected to be tetraquark states \cite{CP04,BG04}
from the consideration of their mass, narrow decay width and decay mode.
As the unusual features of X(3872), 
its mass is rather close to the threshold of D$^0(c\bar{u})$ and $\bar{\rm D}^{0*}(u\bar{c})$, 
and its decay width is very narrow as $\Gamma <$ 2.3MeV (90 \% C.L.). 
These facts seem to indicate that X(3872) is a tetraquark \cite{CP04}
or a molecular state of D$^0(c\bar{u})$ and $\bar{\rm D}^{0*}(u\bar{c})$ 
rather than an excited state of a $c\bar c$ system \cite{BG04}.
Also, D$_{\rm s0}^+$(2317) is conjectured to be not a simple meson of $c\bar{s}$ 
but a tetraquark from its small mass and its narrow decay width.
In fact, simple quark-model assignment for D$_{\rm s0}^+$(2317) is $c\bar s(^3P_0)$,
but almost all theoretical predictions for P-wave $c\bar s$ states 
fail to reproduce its small mass \cite{BG04,LW00}.
Furthermore, there is a puzzle in the decay mode of D$_{\rm s0}^+$(2317): 
CLEO collaboration \cite{Ds} shows that  
the dominant decay mode is D$_{\rm s}^{+} \pi^0$ and 
its contribution is much larger than the radiative decay as 
\vspace{-0.2cm} 
\begin{eqnarray}
\frac{\Gamma({\rm D}_{\rm s0}^+ \rightarrow {\rm D}_{\rm s}^{*+} \gamma)}
     {\Gamma({\rm D}_{\rm s0}^+ \rightarrow {\rm D}_{\rm s}^{+} \pi^0)} < 0.059.
\end{eqnarray}
If D$_{\rm s0}^+$(2317) is isoscalar ($I$=0) such as $c\bar s$, 
this decay pattern implies that the main decay mode is isospin breaking process 
of $O(e^4)$ in QED, which is rather anomalous.
To explain this decay pattern, Terasaki proposed $I=1$ possibility of D$_{\rm s0}^+$(2317) \cite{T03}, 
which means the tetraquark picture for D$_{\rm s0}^+$(2317) and leads to 
D$_{\rm s0}^{++}$ as the isospin partner. 

Then, we examine manifestly exotic tetraquark D$_{\rm s0}^{++} (cu\bar s\bar d)$ 
with $C=+1$, $S=+1$ and $I_3=+1$ in anisotropic quenched lattice QCD.
In this paper, we mainly report the idealized SU(4)$_f$ case.

\vspace{-0.1cm}

\subsection{Tetraquarks in the light-quark sector}

There are five $0^{++}$ isoscalar mesons below 2GeV: 
$f_0$(400-1200), 
$f_0$(980), 
$f_0$(1370), 
$f_0$(1500) and 
$f_0$(1710). 
Among them, $f_0$(1370) is considered as the lowest $q\bar q$ scalar meson in the quark model \cite{PDG}.
For, in the quark model, the lowest $q\bar q$ scalar meson is$\ ^3P_0$, 
and therefore it turns to be rather heavy.
[$f_0$(1500) and $f_0$(1710) are expected to be the lowest scalar glueball or an $s\bar s$ scalar meson.]
Then, what are the two light scalar mesons, $f_0$(400-1200) and $f_0$(980)? 
This is the ``scalar meson puzzle", which is unsolved even at present. 
As the possible answer, in 1977, Jaffe proposed tetraquark ($qq\bar q\bar q$) assignment 
for low-lying scalar mesons such as $f_0$(980) and $a_0$(980) \cite{J77}.

Then, we examine the scalar tetraquark $f_0(ud\bar u\bar d)$ in the light-quark sector 
in anisotropic quenched lattice QCD.
In this paper, we only consider connected diagrams at the quenched level. 
Note here that, if only the connected diagram is taken account at the quenched level, 
the tetraquark $f_0(ud\bar u\bar d)$ is identical with 
D$_{\rm s0}^{++} (cu\bar s \bar d)$ in the idealized SU(4)$_f$ case.

\vspace{-0.1cm}

\section{Lattice QCD}

\vspace{-0.1cm}

We investigate the manifestly exotic tetraquark D$_{\rm s0}^{++}(cu\bar s \bar d)$ 
and the scalar tetraquark $f_0(ud\bar u\bar d)$ in anisotropic quenched lattice QCD 
mainly in the idealized SU(4)$_f$ case.

\subsection{Diquark-antidiquark-type interpolating field operator}

For the tetraquark D$_{\rm s0}^{++}(cu\bar s\bar d)$ with spin $J=0$ and isospin $I=1$, 
we adopt the diquark-antidiquark-type (non-two-meson-type) interpolating field,
\begin{eqnarray}
O \equiv \epsilon_{abc}\epsilon_{ade}
(c^T_bC\gamma_5 u_c)(\bar s_dC\gamma_5 \bar d^T_e),
\label{DDbar}
\end{eqnarray}
to extract the 4Q resonance rather than two-meson scattering states.
Here, the roman indices denotes color indices, 
and $C \equiv \gamma_4\gamma_2$ the charge conjugation matrix. 
In this diquark-antidiquark-type operator, 
the overlap with two-meson scattering states is expected to be small.
The energy of the low-lying state is extracted from 
the temporal 4Q correlator 
$G(t)\equiv \frac{1}{V}\sum_{\vec x}\langle O(t,\vec x)
O^\dagger (0,\vec 0)\rangle$,  
where the total momentum of the 4Q system is projected to be zero.
To reduce highly-excited-state components, 
we use a spatially-extended source of Gaussian-type with $\rho$=0.4fm 
in the Coulomb gauge \cite{IDIOOS05,MOU01}. 
We impose the Dirichlet boundary condition 
in the temporal direction \cite{TUOK05}.
We analyze the effective mass $m_{\rm eff}(t)$ obtained from the 4Q correlator $G(t)$ as 
$
m_{\rm eff}(t)\equiv {\rm ln}\{G(t)/G(t+1)\}.
$

\vspace{-0.1cm}

\subsection{Anisotropic lattice QCD}

To get detailed information on the temporal behavior of the 4Q correlator $G(t)$, 
we adopt anisotropic lattice QCD with 
the standard plaquette action and the $O(a)$-improved Wilson (clover) fermion 
at $\beta \equiv 2N_c/g^2=5.75$ on $12^3 \times 96$ with renormalized anisotropy $a_s/a_t=4$ \cite{IDIOOS05,MOU01}. 
This is anisotropic version of the Fermilab action \cite{EKM97}.
The scale is set by the Sommer scale $r_0^{-1}=395{\rm MeV}$, which leads to the spatial/temporal lattice spacing as 
$a_s^{-1}\simeq 1.10{\rm GeV}$ (i.e., $a_s\simeq$ 0.18fm) and $a_t^{-1}\simeq 4.40{\rm GeV}$ (i.e., $a_t\simeq$ 0.045fm).
The spatial lattice size is $L =12 a_s \simeq 2.15{\rm fm}$. 
We use 1827 gauge configurations, which are picked up every 500 sweeps after the thermalization of 10,000 sweeps. 
On the happing parameter, we take $\kappa$=0.1210-0.1240 for light quarks and 
$\kappa$=0.1120 for the real charm-quark mass.
We summarize the lattice parameters and related quantities in Table~1.

\vspace{-0.1cm}

\begin{table}[ht]
\begin{center}
\label{parameters}
\caption{The lattice parameters and related quantities. 
They are almost the same as in Ref.\cite{IDIOOS05}.}
\vspace{-0.2cm}
\begin{tabular}{ccccccccc}
\hline
\hline
$\beta$  & lattice size & $a_s^{-1}$ & $a_t^{-1}$ &$\gamma_G$ & $u_s$ & $u_t$ & $\kappa$ 
\\
\hline
5.75 & $12^3 \times 96$& 1.10GeV & 4.40GeV&3.2552 &0.7620 &0.9871 & 0.1210(10)0.1240, 0.1120\\
\hline
\hline
\end{tabular}
\end{center}
\vspace{-0.65cm}
\end{table}

\vspace{-0.2cm}

\subsection{Hybrid Boundary Condition (HBC) method}

To extract low-lying 4Q resonances, 
we use the ``hybrid boundary condition" (HBC) \cite{IDIOOS05} 
where we impose the the anti-periodic boundary condition for quarks ($c$, $u$) and 
the periodic boundary condition for antiquarks ($\bar s, \bar d$), as shown in Table~2.
By applying the HBC on a finite lattice with $L^3$, the two-meson threshold is raised up, 
while the mass of a compact 4Q resonance is almost unchanged. 
Then, the 4Q resonance may become visible as a low-lying state in HBC, if it exists.

\vspace{-0.3cm}

\begin{table}[hb]
\newcommand{\cc}[1]{\multicolumn{1}{c}{#1}}
\caption{
The hybrid boundary condition (HBC) to raise up the threshold of 
two-meson scattering states.}
\vspace{-0.7cm}
\begin{center}
{\large
\begin{tabular}{llllll} \hline \hline
  & $c$, $u$ & $\bar s$, $\bar d$  & $q\bar q$-meson & 2-meson threshold & 
tetraquark ($qq\bar q\bar q$)\\
\hline 
PBC & periodic & periodic & periodic & $m_1+m_2$ & periodic \\
HBC & anti-periodic & periodic & anti-periodic & 
$\sum_{k=1,2}\sqrt{m_k^2+\vec p_{\rm min}^2}$ & periodic  \\
\hline\hline
\end{tabular}
}
\end{center}
\vspace{-0.5cm}
\end{table}

For a compact 4Q quasi-bound state of $cu \bar s \bar d$, 
since it contains even number of quarks, 
it obeys periodic boundary condition (PBC) in HBC, 
and therefore its energy in HBC is almost the same as that in PBC \cite{IDIOOS05}.  
For a two-meson scattering state, 
both mesons have non-zero relative momentum 
$\vec p_{\min}=(\pm\frac{\pi}{L},\pm\frac{\pi}{L},\pm\frac{\pi}{L})$, i.e.,
$|\vec p_{\rm min}|=\sqrt{3}\pi/L$ in HBC, 
while they can take zero relative momentum $\vec p_{\rm min}=0$ in PBC.
In fact, the two-meson threshold $m_1+m_2$ ($m_1$, $m_2$: meson masses) in PBC 
is raised up in HBC as 
$E_{\rm th} \simeq \sum_{k=1,2}\sqrt{m_k^2+\vec p_{\rm min}^2}$ 
with $|\vec p_{\rm min}| =\sqrt{3} \pi/L \simeq 0.5 {\rm GeV}$ for 
$L \simeq 2.15 {\rm fm}$.

\vspace{-0.2cm}

\section{Lattice QCD results for tetraquarks}

\vspace{-0.1cm}

First, for comparison, we calculate the $q\bar q$ scalar meson mass in anisotropic quenched lattice QCD.
Here, we only consider connected diagrams. 
We show in Fig.1 the lowest $q\bar q$ scalar meson mass plotted against $m_\pi^2$.
Our quenched lattice QCD indicates that the lightest $q\bar q$ scalar meson 
has a large mass about 1.37 GeV after the chiral extrapolation. 
Thus, the lightest $q\bar q$ scalar meson corresponds to $f_0(1370)$ or $a_0(1450)$, which 
is consistent with the quark-model assignment for scalar mesons.
[Note that, if disconnected diagrams are dropped off, 
isoscalar $q\bar q$ mesons degenerate isovector $q\bar q$ mesons in quenched QCD, 
e.g., $m_\rho=m_\omega$ and $m(f_0)=m(a_0)$.]

\begin{figure}
\begin{center}
\begin{minipage}{7cm}
\includegraphics[width=6cm]{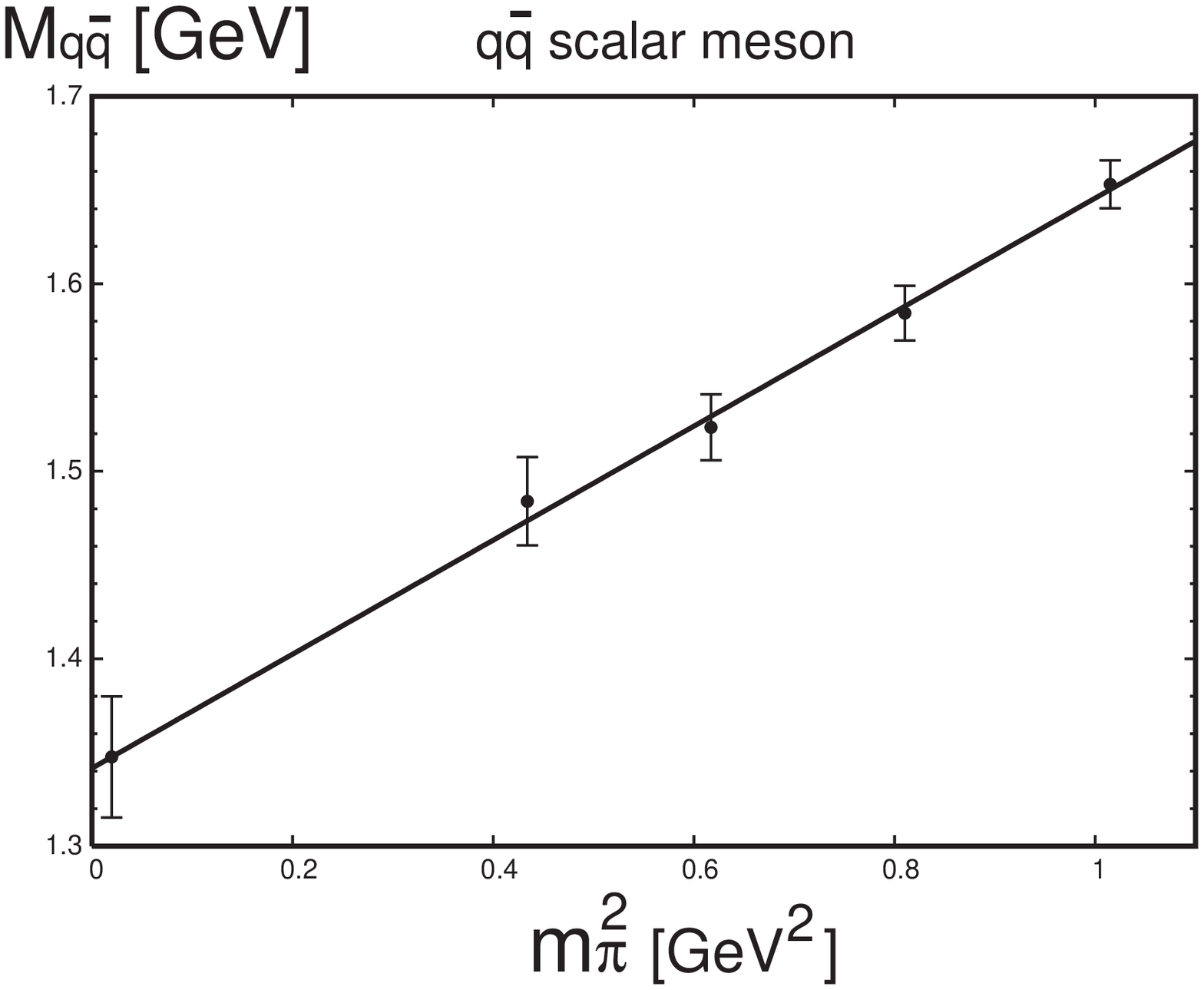}
\vspace{-0.5cm}
\caption{The lowest mass $M_{q\bar q}$ of the $q\bar q$ scalar meson plotted against $m_\pi^2$.
The $q\bar q$ scalar meson mass $M_{q\bar q}$ is extracted from the connected diagram 
in quenched lattice QCD.
}
\label{fig1}
\end{minipage}
\hspace{0.2cm}
\begin{minipage}{7cm}
\includegraphics[width=6.2cm]{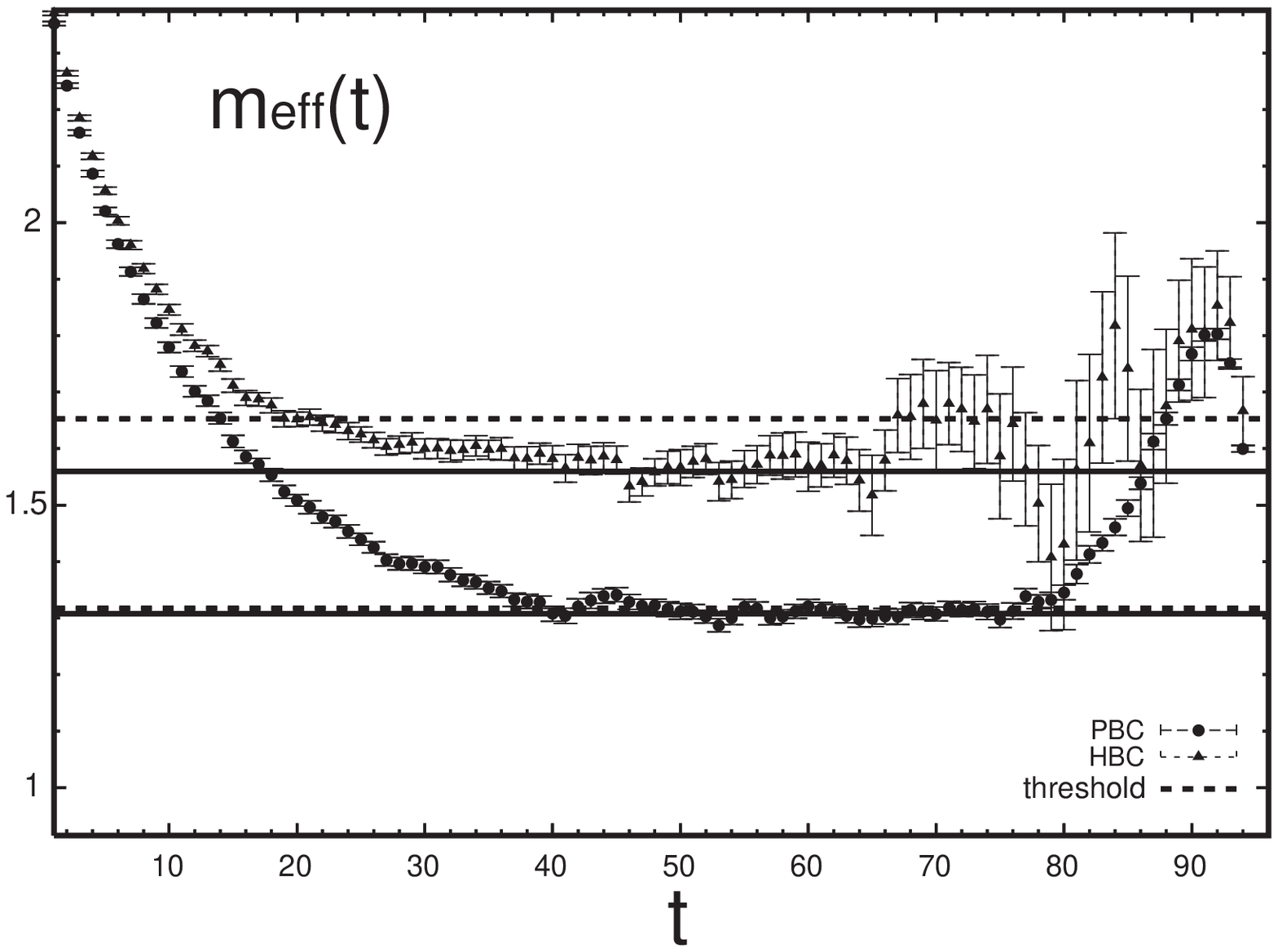}
\vspace{-0.2cm}
\caption{
The effective mass plot of 4Q system in PBC and HBC.
The solid lines denote the plateau of $m_{\rm eff}(t)$, 
and the the dashed lines the two-meson threshold.
The Dirichlet boundary condition is imposed in the temporal direction. 
}
\label{fig2}
\end{minipage}
\hspace{0.2cm}
\end{center}
\vspace{-0.75cm}
\end{figure}

\begin{figure}[ht]
\begin{center}
\begin{minipage}{7cm}
\includegraphics[width=5.6cm]{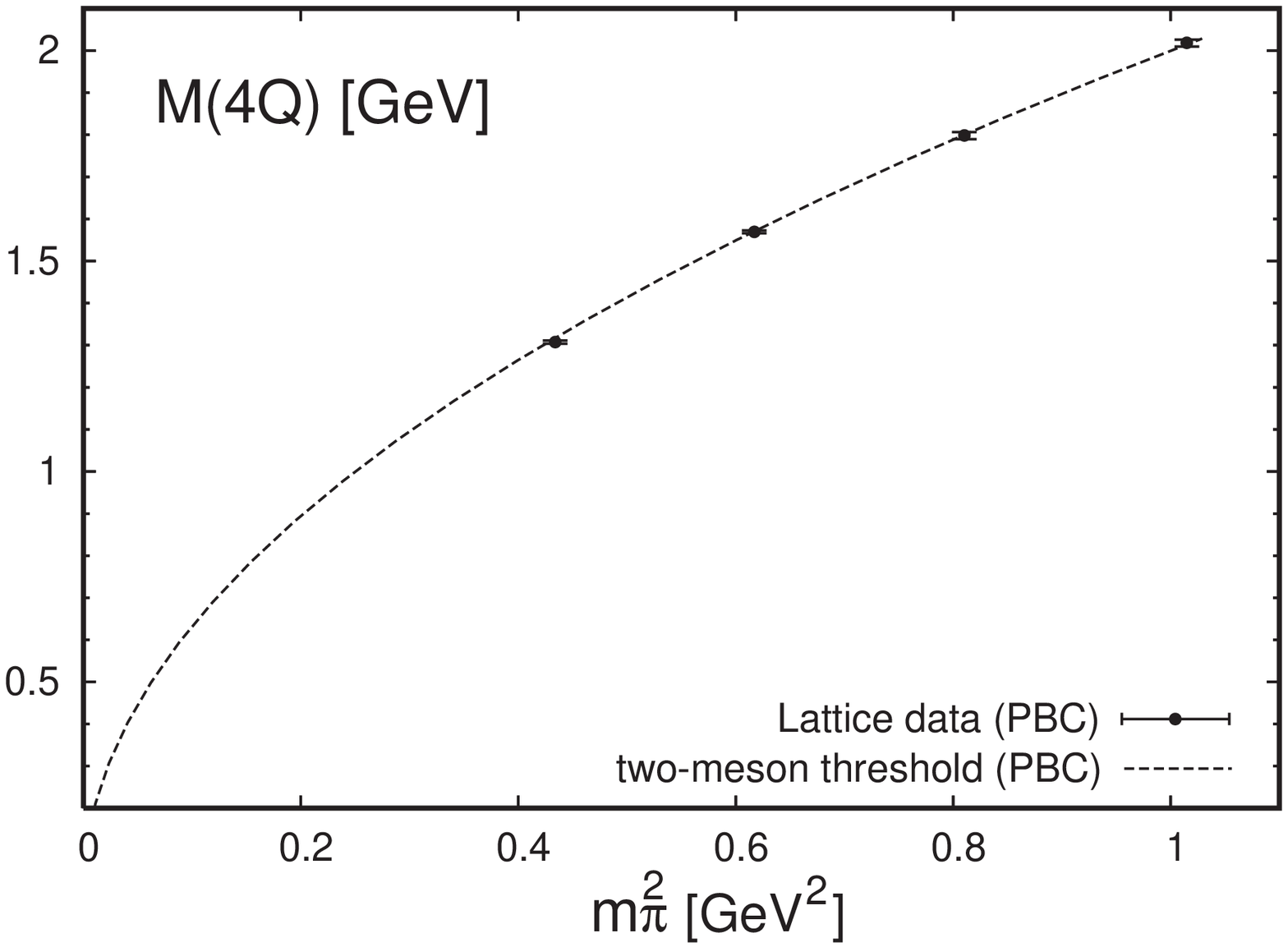}
\vspace{-0.2cm}

\caption{
The lowest energy $M$(4Q) of the 4Q system in PBC plotted against $m_\pi^2$.
The dotted curve denotes the two-meson threshold of $2m_\pi$.  
}
\label{fig3}
\end{minipage}
\hspace{0.2cm}
\begin{minipage}{7cm}
\includegraphics[width=5.6cm]{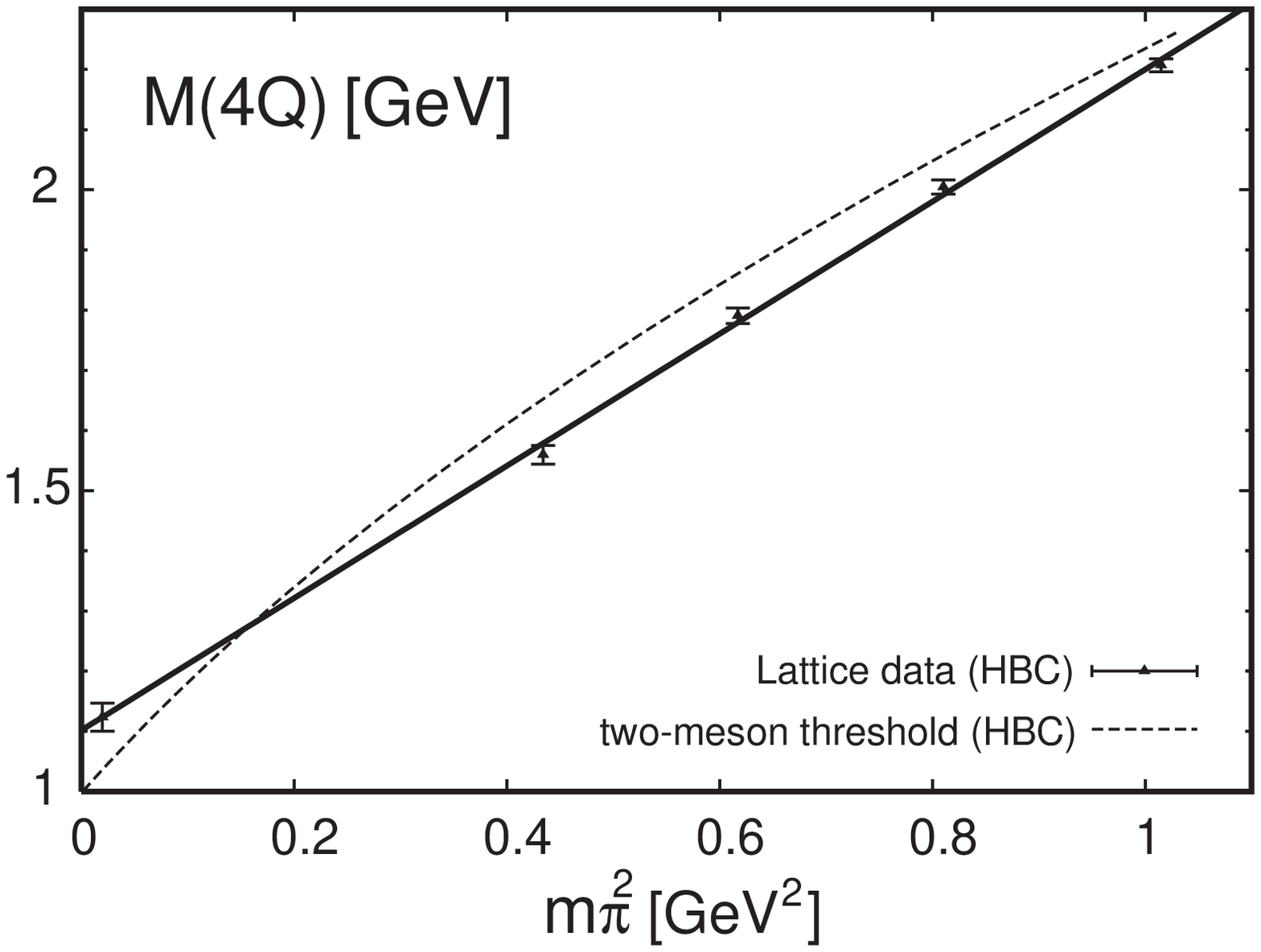}
\vspace{-0.2cm}

\caption{
The lowest energy $M$(4Q) of the 4Q system in HBC v.s. $m_\pi^2$.
The dotted curve is the two-meson threshold of 
$\sqrt{m_\pi^2+3\pi^2/L^2}$ in HBC.  
}
\label{fig4}
\end{minipage}
\end{center}
\vspace{-0.75cm}
\end{figure}

Second, we investigate the 4Q system of 
$cu \bar s \bar d$ with various quark masses in PBC and in HBC.
Here, we show the idealized SU(4)$_f$ case, 
where D$_{\rm s}^{++}(cu \bar s \bar d)$ is identical with 
$f_0(ud \bar u \bar d)$ as far as only connected diagrams is considered.
Figure 2 shows the effective mass $m_{\rm eff}(t)$ of 
$cu\bar s \bar d$ (or $ud \bar u \bar d$) in PBC and in HBC for $\kappa$=0.1240. 
In PBC, the plateau of $m_{\rm eff}(t)$ agrees with the two-pseudoscalar-meson threshold, 
i.e., 2$m_\pi$.
As shown in Fig.3, the lowest energy of the 4Q system in PBC plotted is 
just a $\pi$-$\pi$ scattering state, as is expected.
[Note that this lowest 4Q level is raised up in HBC, while 
a compact 4Q resonance does is almost unchanged in HBC.]

Third, we analyze the lowest 4Q state in HBC.
In HBC, there appears a plateau below the $\pi$-$\pi$ threshold by about 100MeV as shown in Fig.3.
Figure 4 shows the lowest energy of the 4Q system in HBC plotted against $m_\pi^2$.
Note that its chiral behavior is largely different from the $\pi$-$\pi$ scattering state.
Thus, this 4Q state is considered to be a nontrivial 4Q resonance instead of a two-meson scattering state.
After the simple linear chiral extrapolation, 
the mass of the 4Q state is estimated to be about 1.1GeV, 
which seems to correspond to $f_0$(980) or $a_0(980)$ in the light-quark scalar sector.
This also indicates that the manifestly exotic tetraquark D$_{\rm s0}^{++}(cu\bar s \bar d)$ 
would exist around 1GeV in the idealized SU(4)$_f$ chiral limit.

\vspace{-0.3cm}

\section{MEM analysis for point-point 4Q correlation}

\vspace{-0.2cm}

Finally, using the maximum entropy method (MEM) \cite{AHN01}, 
we extract the spectral function $A(\omega)$ 
from the temporal correlator $G(t)$, which satisfies 
\vspace{-0.2cm}
\begin{eqnarray}
G(\tau)= \int_0^\infty d\omega K(\tau,\omega)A(\omega), 
~~~~~~K(\tau,\omega)\equiv \frac{e^{-\tau\omega}+e^{-(\beta-\tau)\omega}}{1-e^{-\beta\omega}},
~~~~~\beta \equiv N_ta_t.
\label{disp}
\end{eqnarray}
From the lattice data of the 4Q correlator $G(t)$, we obtain the spectral function $A(\omega)$ of the 4Q system with MEM.
At present, we perform the MEM analysis for point-source point-sink 4Q correlator $G(t)$ 
both in PBC and in HBC, although this correlator includes large excited-state contamination.
For the point-point correlator $G(t)$, the lowest-order default function $m(\omega)$ 
is calculated as 
\vspace{-0.1cm}
\begin{eqnarray}
m(\omega)=\frac{4N_c}{2^8 \Gamma(5)\Gamma(6)\pi^6}\omega^8.
\label{default}
\end{eqnarray}
Unlike the previous section, we here use $\kappa=0.1240$ for $u$, $d$, $s$-quarks, 
and $\kappa=0.1120$ for $c$-quark, 
and impose the periodic boundary condition in temporal direction.
Figures 5 and 6 show the MEM analysis for the 4Q correlator $G(t)$ and 
the obtained spectral function $A(\omega)$ for the 4Q system.

\vspace{-0.3cm}
\begin{figure}[hb]
\begin{center}
\begin{minipage}{7cm}
\includegraphics[width=6cm]{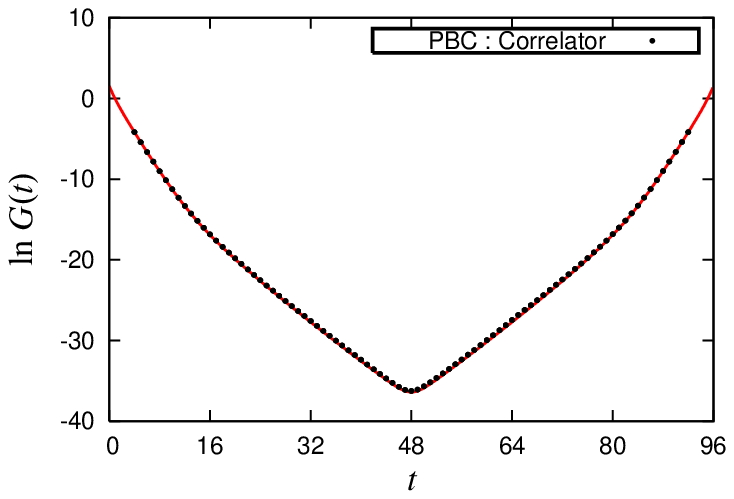}
\vspace{-0.5cm}
\caption{
The 4Q corelator of point-source point-sink correlator $G(t)$ in PBC.
The symbols denote the lattice data. The solid curve is 
the temporal correlator reconstructed by $A(\omega)$ in Fig.6.
}
\label{fig5}
\end{minipage}
\hspace{0.2cm}
\begin{minipage}{7cm}
\includegraphics[width=6.2cm]{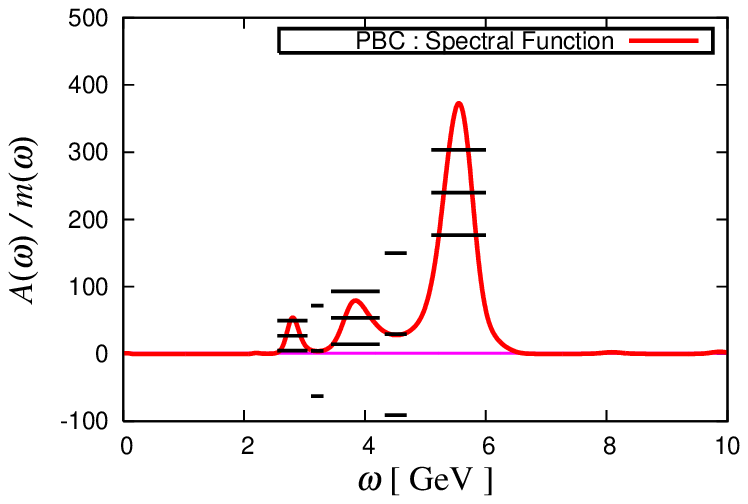}
\vspace{-0.2cm}

\caption{
The spectral function $A(\omega)$ 
obtained with MEM 
from the point-source point-sink 4Q correlator in PBC. 
}
\label{fig6}
\end{minipage}
\end{center}
\vspace{-0.6cm}
\end{figure}

\noindent
The lattice data of 4Q correlator $G(t)$ 
can be well reconstructed with the spectral function $A(\omega)$ obtained by MEM 
through Eq.(\ref{disp}).
[In this point-point correlator $G(t)$, highly-excited contamination is so large that 
low-lying structure of $A(\omega)$ cannot be seen clearly.
To clarify the low-lying structure of the 4Q spectrum, 
we are performing the MEM analysis for extended-source extended-sink 4Q correlators, 
where excited-state components are reduced.]

\vspace{-0.1cm}

\section{Summary and conclusions}

\vspace{-0.1cm}

We have studied the manifestly exotic tetraquark
D$_{\rm s0}^{++}(cu\bar s \bar d)$ and 
the scalar tetraquark $f_0(ud \bar u \bar d)$ 
in SU(3)$_c$ anisotropic quenched lattice QCD  
mainly in the idealized SU(4)$_f$ case.
First, we have found that the lowest $q\bar q$ scalar meson 
has a large mass of about 1.37GeV after chiral extrapolation, 
which corresponds to $f_0(1370)$.
Second, we have found that the lowest 4Q state 
is just a scattering state of two pseudoscalar mesons. 
Third, 
using the Hybrid Boundary Condition (HBC) method, 
we have found a nontrivial 4Q resonance state 
whose chiral behavior is different from a two-meson scattering state.
The scalar tetraquark $f_0(ud\bar u\bar d)$ is found to have the mass of 
about 1.1GeV after chiral extrapolation, and seems to correspond to $f_0(980)$.
Then, the manifestly exotic tetraquark D$_{\rm s0}^{++}(cu\bar s \bar d)$ 
is expected to exist around 1GeV in the idealized SU(4)$_f$ chiral limit.

\end{document}